\documentstyle[preprint,aps,epsfig]{revtex}

\newcommand{\be}{\begin{eqnarray}}
\newcommand{\ee}{\end{eqnarray}}

\newcommand{\k}{\kappa}
\newcommand{\e}{\epsilon}

\begin{document}


\preprint{
YUMS 97--17
}

\title{ 
Implications of the recent CERN LEP data on nonuniversal interactions\\
with the precision electroweak tests
}
\vspace{0.65in}

\author{
Kang Young ~Lee\thanks{kylee@chep5.kaist.ac.kr}
\footnote{Present address: Phys. Dept., KAIST, Taejon 305-701, Korea}
}
\vspace{.5in}

\address{
Department of Physics, Yonsei University, Seoul 120--749, Korea}

\date{\today}
\maketitle

\begin{abstract}
\\
We explore the nonuniversal interaction effects
in terms of the precision variables epsilons
with the recent LEP data reported by the Electroweak Working Group.
The epsilon variables with the nonuniversal interactions
are calculated and constrained by the experimental ellipses
in the $\epsilon_1$--$\epsilon_b$, $\epsilon_2$--$\epsilon_b$,
and $\epsilon_3$--$\epsilon_b$ planes.
We find that the new data enables us to make 
a stringent test on the correction to $Z \to b \bar{b}$ vertex.
The $\epsilon_b$ variable is sensitive to the $Z b \bar b$
couplings and thus plays a major role to give constraints on 
the nonuniversal interaction effects.
Upon imposing the new data on $\epsilon_b$,
we have the allowed range of the model parameter
$\kappa_L = 0.0063 \pm 0.0030$ at 1-$\sigma$ level with $m_t = 175$ GeV.
Along with the minimal contact term, we predict the new physics scale
$\Lambda \sim$ 1.6 TeV.
By combining the experimental results 
from all planes
we obtain the allowed range of $\kappa_L$ : $0.003 < \kappa_L < 0.010$
at 95 \% C.L..

\end{abstract}

\pacs{ }

\narrowtext


Since the top quark was observed and its mass has been measured
from the Fermilab $p \bar p$ collider Tevatron,
the influence of the large value of the top quark mass 
on the $Z \to b \bar b$ vertex is drawing much attention.
In the year of 1995, the value of 
$R_b \equiv \Gamma(Z \to b \bar b)/\Gamma(Z \to \mbox{hadrons})$
was reported to be actually more than three standard deviations 
higher from the Standard Model (SM) prediction with the heavy top
and it stimulated many theoretical and experimental efforts.
The most recent data from ALEPH \cite{aleph}, DELPHI \cite{delphi}, 
OPAL \cite{opal} and SLD \cite{sld},
however, come closer to the SM prediction.
Following the LEP electroweak working group report \cite{LEPEWWG},
the average of 1996 data (LEP+SLC) is $R_b = 0.2178 \pm 0.0011$,
which is about 1.8 standard deviations higher than the SM prediction.
In spite of the experimental improvement of the situation,
much interest is taken in the $R_b$ problem as ever
since the discrepancy between the measured value and the SM
predicted value still exists. 
Furthermore recent data have different systamatics 
from those obtained until 1995 and it is not clear
whether it is appropriate to discard old measurements
from the world average.
Therefore it is still interesting to consider the possibility
of the new physics beyond the SM which affect 
the $Z \to b \bar{b}$ vertex. 

The nonuniversal interaction acting only on the third generation
can be an attractive candidate for the new physics, 
at least in the viewpoint of the effective theory,
since we favor that the SM predictions for other flavours 
should not be much disrupted by the new physics.
Such models are mainly motivated by the idea that mass of the top quark 
turns out to be of order of the weak scale
and so the top quark could be responsible for the electroweak
symmetry breaking.
The top quark condensation may be formed
if we introduce a new gauge interaction and the bound state
$\langle \bar{t} t \rangle$ would play the role of the Goldstone
bosons for the electroweak symmetry breaking 
instead of the elementary scalar.

In this paper 
we attempt to constrain the nonuniversal interactions 
in terms of the precision variables $\epsilon$'s.
We consider the general approach to introduce
the nonuniversal corrections to the $Z \to b \bar b$ vertex 
in a model independent way.
Since the anomalous nonuniversal interaction terms should be
SU(2)$_L \times$U(1)$_Y$ invariant, $b$--quark also interacts
with $t$--quark via involving the left--handed doublet interactions.
This anomalous interaction results in the additional contributions
to the $Z \to b \bar b$ vertex.
We parametrize the nonuniversal interaction effects in the 
$Z \to b \bar b$ vertex by the shift of the tree level SM 
couplings of the neutral currents $g_{L,R}$.
We let the effective couplings $g_{L,R}^{\mbox{\footnotesize{eff}}}$:
\be
g_{L,R}^{\mbox{\footnotesize{eff}}}
=g_{L,R} (1 + \k_{L,R}) ~~,
\ee
where
\be
g_L = -\frac{1}{2} + \frac{1}{3} \sin^2 \theta_W,~~~
g_R =  \frac{1}{3} \sin^2 \theta_W ~~.
\nonumber
\ee

In order to parametrize the new physics effects, we use the precision
variables $\epsilon_i$'s introduced by Altarelli et al.
\cite{altarelli1} here.
We calculated the epsilons for the updated analysis
of the precision electroweak tests of the supergravity models 
by taking into account the new LEP data presented at the 28th ICHEP
(1996, Poland)
to obtain more accurate experimental values of $\epsilon_{1,2,3,b}$
in the ref. \cite{parklee}.
Out of the epsilon variables, $\epsilon_b$ has been of particular interest
because it encodes the loop corrections to the $Z \to b \bar b$ vertex 
and is relevant for our aim.
Our new physics corrections $\kappa_{L,R}$ are generically due to 
the sum of the bubble diagram of the top quark in this type of model.
We calculate $\epsilon_{1,2,3,b}$ with nonuniversal corrections and
constrain them by the results of the experimental data.


In the original work \cite{altarelli2}, the variables 
$\e_1$, $\e_2$ and $\e_3$ were defined from the basic observables,
the mass ratio of $W$ and $Z$ bosons $m_W/m_Z$, the leptonic
width $\Gamma_l$ and the forward--backward asymmetry for
charged leptons $A_{FB}^l$.
These observables are all defined at the $Z$--peak,
precisely measured and free from important
hadronic effects like $\alpha_s(m_Z)$ or the $Z \to b \bar b$
vertex.
In terms of these observables, $\e_1$, $\e_2$ and $\e_3$,
we have the virtue that the most interesting physical results are 
already obtained at a completely model independent level
without assumptions like the dominance of vacuum polarisation diagrams.

Because of the large $m_t$--dependent SM corrections to 
the $Z \to b \bar b$ vertex, however, 
the $\epsilon_i$'s and $\Gamma_b$ can only be 
correlated for a given value of $m_t$.
In order to overcome this limitation, Altarelli et al. \cite{altarelli1}
added a new parameter, $\epsilon_b$, which encodes the $m_t$--dependent
corrections to $Z \to b \bar b$ vertex and slightly modified 
other $\epsilon_i$'s.
Hence the four $\epsilon_i$'s are defined from an enlarged set
of basic observables $m_W/m_Z$, $\Gamma_l$, $A_{FB}^l$ and $\Gamma_b$
without need of specifying $m_t$.
Consequently the $m_t$--dependence for all observables
via loops come out through the $\epsilon_i$'s.
We work with this extended scheme here because we are interested in
the corrections to $Z \to b \bar b$ vertex.


Since the nonuniversal corrections affect only on the
$Z \to b \bar b$ vertex among the basic observables,
we focus on the variable $\epsilon_b$ here.
The correction to $g_R$ does not affect $\Gamma_b$ significantly
because $g_L \gg g_R$ in eq. (1), 
and we neglect the effect of $\kappa_R$ hereafter.
We calculate the epsilon variables with the nonuniversal corrections
using the ZFITTER \cite{ZFITTER}.
Fig. 1 shows the $\kappa_L$--dependence of the $\epsilon_b$ variable
for $m_t = 175$ GeV.
We know that $\epsilon_b$ is insensitive to the variation 
of the Higgs mass $m_H$.
The range of $\kappa_L$ corresponding to the 1-$\sigma$ of the
experimental data is given by
\be
\kappa_L = 0.0033 \sim 0.0093~~.
\ee

The epsilon variables are obtained in the ref. \cite{parklee}
from the recent LEP data given in table I 
reported by the LEP Electroweak Working Group \cite{LEPEWWG}:
\be
\epsilon_1 &=& (4.0 \pm 1.2) \times 10^{-3}
\nonumber \\
\epsilon_2 &=& (-4.3 \pm 1.7) \times 10^{-3}
\nonumber \\
\epsilon_3 &=& (2.3 \pm 1.7) \times 10^{-3}
\nonumber \\
\epsilon_b &=& (-1.5 \pm 2.5) \times 10^{-3}~~.
\ee
Note that the lepton universality assumption
is assumed for the values of $\Gamma_l$ and $A_{FB}^l$.
Besides in the $\epsilon_1-\epsilon_b$ plane,
we attempt to constrain the model by the experimental ellipses
in the $\epsilon_2-\epsilon_b$ and $\epsilon_3-\epsilon_b$ planes
here.
In Fig. 2, the experimental ellipses for 1-$\sigma$ level and
90\%, 95\% confidence level are given in the
$\epsilon_1-\epsilon_b$ plane (a), in the $\epsilon_2-\epsilon_b$ 
plane (b) and in the $\epsilon_3-\epsilon_b$ plane (c)
with our model predictions for varying the parameter
$\kappa_L$ and the Higgs mass $m_H$.
We find the SM results ($\kappa_L =0$) deviate even from
the 95 \% C.L. ellipses for all of three cases
and that the nonuniversal corrections improve the situations in general.
$\epsilon_1$ and $\epsilon_2$ favor the heavy Higgs and
$\epsilon_3$ favors the light Higgs mass $\sim$ 100 GeV.
As the more precise value of the $W$ boson mass is reported,
$\epsilon_2$ variable can also provide a stringent test 
for the theoretical predictions.
Here, we used the value of the $W$ boson mass fitted to LEP data alone
by LEP Electroweak Working Group \cite{LEPEWWG}.
As its precise measurement will be performed at LEP II, 
we can expect the progress in $\epsilon_2$ analysis.
We find that $\epsilon_3$ demands the new physics most strongly
among them and that the Higgs mass get the upper bound
$m_{_H} \lesssim 300$ GeV at 95 \% C. L..
Combining the experimental ellipses conditions
on the $\epsilon_1$--$\epsilon_b$, $\epsilon_2$--$\epsilon_b$,
and $\epsilon_3$--$\epsilon_b$ planes,
we obtained the range of allowed values of $\kappa_L$ :
$0.003 < \kappa_L < 0.010$ at 95 \% C. L. with the Higgs mass
$m_{_H} = 100 \sim 300$ GeV.
The heavier the Higgs, the narrower the aloowed region.
At 90 \% C.L., we obtain extremely small region :
$\kappa_L \sim 0.007$ and $m_{_H} \sim 120$ GeV.
In our analysis, we use the values $\alpha_s(m_Z) = 0.118$
and $\alpha(m_Z) = 1/128.87$.

The mass of the top quark is being measured more precisely
by the CDF and D0 collaborations at Tevatron.
As stated before, we use 175 GeV as input value of $m_t$ 
in Fig. 1 and Fig. 2, which is the central value 
of the recent CDF and D0 report \cite{top}.
The $Z \to b \bar b$ vertex is, however, affected much
by the change of $m_t$ and we present the $m_t$--dependence
of $\epsilon_b$ in Fig. 3.
The value of $m_t$ is varied from 170 to 180 geV.
If $m_t = 170$ GeV, the allowed range of $\kappa_L$ is given by
$\kappa_L = 0.0032 \sim 0.0088$ and if $m_t = 180$ GeV,
$\kappa_L = 0.0039 \sim 0.0099$ at 1-$\sigma$ level.
We find that the value of $\epsilon_b$ is not significantly changed
in this range of $m_t$.
The major features of the constraints from $\epsilon_{1,2,3,b}$
for the nonuniversal interactions are summarized in table II.



In this work we explored the nonuniversal interaction effects
on the $Z \to b \bar b$ vertex in terms of the $\epsilon$ variables
using the recent experimental data.
We did not explicitly describe the parameter $\kappa_L$
by specific physical quantity here since we take a model--independent
approach.
Various models which can give the effective Lagrangian
for the $Z \to b \bar b$ vertex
\be
{\cal L}_{eff} \sim Z^{\mu} (\bar{b} \gamma_{\mu}
      (g_V^{eff} + g_A^{eff} \gamma_5) b)~~
\ee
have been considered by several authors \cite{bamert,kkll,hill1,zhang}.
One of the most appropriate type of models is the top condensation
idea in which the third generation has their own gauge interaction.
In general we have the several contact terms which are $d>4$
at a high energy scale in those models.
We find a general list of contact terms 
in ref. \cite{buchmuller,hill2}.
As the minimal contents of the model, left--handed SU(2) doublet for
the third generation and the right--handed singlet $t_R$ are
coupled in a new gauge interaction.
We can write a relevant term of the effective Lagrangian as
\be
{\cal L}_{eff} = - \frac{1}{\Lambda^2} \bar{b} \gamma_{\mu} b
		   \bar{t} \gamma^{\mu} (g_V-g_A \gamma_5) t
		   + ...
\ee
where $g_V$, $g_A$ are model parameters.
Then the effective contribution to $Z \to b \bar b$ vertex,
$\kappa_L$, is generated via the top quark loops 
thus we obtain the relation,
\be
\kappa_L = \frac{g_A}{g_L} \frac{N_c}{8 \pi^2} \frac{m_t^2}{\Lambda^2}
	   \ln \left( \frac{\Lambda^2}{m_t^2} \right)~~.
\ee
The central value from the experimental data $\epsilon_b = -1.5$
leads to the value of parameter $\kappa_L \sim 0.0063$
and yields the new physics scale $\Lambda \sim$ 1.6 TeV.
We find that the new physics scale is rather low and it enables us to
avoid the extra fine--tuning of the new gauge couplings for the 
hierarchy between $m_t$ and $\Lambda$.
On the other hand, such a four fermion interaction is not enough
for the electroweak symmetry breaking. 
Hill suggested a model in which an separate mechanism 
like extended technicolor is involved to account for 
the observed $W$ and $Z$ masses \cite{topcolor}.
Our analysis is applicable to that model because we pay our attention
to only the influences on the $Z \to b \bar b$ vertex.
With $\kappa_L = 0.0063$ corresponding to the central value from data, 
we calculate the $R_b = 0.2175$,
which agrees with the experimental results from LEP and SLC,
as we expected.
For the $R_b$ from LEP data, we obtain the range of $\kappa_L$:
\be
\kappa_L = 0.0038 \sim 0.0110~~.
\ee
at 1-$\sigma$ level, of which the values are slightly larger than
those from the $\epsilon_b$.

In conclusion, we presented an analysis for the extension of the SM
with the nonuniversal interactions in terms of the precision variables
$\epsilon$'s. 
As a result of the better accuracy of the precision test
with new data from the LEP, the study of the epsilon varibles
provide positive hints for new physics beyond the SM 
and the nonuniversal interactions, at least as an effective theory,
could be a good candidate for the new physics.

\acknowledgements

We would like to thank Prof. Gye T. Park for his directing
us to this subject and valuable comments.
We also thank Dr. Jae Sik Lee for his help for coding
with the ZFITTER.
K. Y. L. is a postdoctoral fellow supported by Korea Research Foundation
(KRF) and Research University Fund of College of Science
at Yonsei University by Ministry of Education (MOE) of Korea.
This work was supported in part by 
the Korean Science and Engineering Foundation (KOSEF).

\begin{table}
\begin{center}
\caption{
The LEP data reported by the LEP Electroweak Working Group
at the 28th ICHEP (1996, Poland).
}
\label{Table1}
\vspace{2cm}
\begin{tabular}{|clc|ccc|}
 & $M_W$ & & & 80.2780  $\pm$  0.0490 GeV& \\
 & $M_Z$ & & & 91.1863  $\pm$  0.0020 GeV& \\
 & $\Gamma_l$ & & & 83.91  $\pm$  0.11 MeV& \\
 & $A_{FB}^l$ & & & 0.0174  $\pm$  0.0010 & \\
 & $\Gamma_b$ & & & 379.9  $\pm$  2.2 MeV& \\
\end{tabular}
\end{center}
\end{table}

\vspace{2cm}

\begin{table}
\begin{center}
\vspace{1cm}
\caption{
The major features of the constraints from $\epsilon_b$
and all ellipses in $\epsilon_1$--$\epsilon_b$, 
$\epsilon_2$--$\epsilon_b$, and $\epsilon_3$--$\epsilon_b$ planes.
for the nonuniversal correction to $Z \to b \bar{b}$ vertex.
}
\vspace{2cm}
\begin{tabular}{ccccccc}
&&&&&&\\
 & $m_t$ &  & $\epsilon_b$ constraints & & combined ellipses constraints & \\
&&&&&&\\
\hline
 & 170 GeV &  & $\kappa_L = 0.0028 \sim 0.0088$ & & & \\
 &         &  & at 1--$\sigma$ level & & & \\
\hline
 & &  & & & $0.004 < \kappa_L < 0.010$ at 95 \% C. L. & \\
 & &  & & & when $m_{_H} \sim 100$ GeV & \\
 & &  & & & $0.003 < \kappa_L < 0.009$ at 95 \% C. L. & \\
 & 175 GeV &  & $\kappa_L = 0.0033 \sim 0.0093$ 
 & & when $m_{_H} \sim 200$ GeV & \\
 & &  & at 1--$\sigma$ level
 & & $0.004 < \kappa_L < 0.008$ at 95 \% C. L. & \\
 & &  & & & when $m_{_H} \sim 300$ GeV & \\
 & &  & & & excluded when $m_{_H} > 300$ GeV & \\
 & &  & & & $\kappa_L \sim 0.007$, $m_{_H} \sim 120$ GeV at 90 \% C. L. & \\
\hline
 & 180 GeV &  & $\kappa_L = 0.0039 \sim 0.0099$ & & & \\
 &         &  & at 1--$\sigma$ level & & & \\
\end{tabular}

\end{center}
\end{table}
\newpage
\vskip 1.0cm
{\large \bf Figure Captions}
\vskip 2.0cm
{\bf Fig. 1}\\
Plot of $\epsilon_b$ in units of $10^{-3}$ 
as a function of the parameter $\kappa_L$
with varying the Higgs mass $m_H$.
The 1-$\sigma$ range obtained from the LEP data is also shown.
\vskip 1.0cm
{\bf Fig. 2}\\
Plot of the model predictions in units of $10^{-3}$ 
with varying the model parameter 
$\kappa_L$ and the Higgs mass $m_H$ in
(a) $\epsilon_1$--$\epsilon_b$ plane,
(b) $\epsilon_2$--$\epsilon_b$ plane and
(c) $\epsilon_3$--$\epsilon_b$ plane.
The experimental ellipses at 1-$\sigma$, 90 \% C.L.
and 95 \% C.L. are also shown.
\vskip 1.0cm
{\bf Fig. 3}\\
Plot of the $m_t$--dependence of the $\epsilon_b$ variable
in units of $10^{-3}$ with varying $\kappa_L$ values.
The 1-$\sigma$ range obtained from the LEP data is
denoted by the dashed line.

\begin{figure}[th]
\centering
\centerline{\epsfig{file=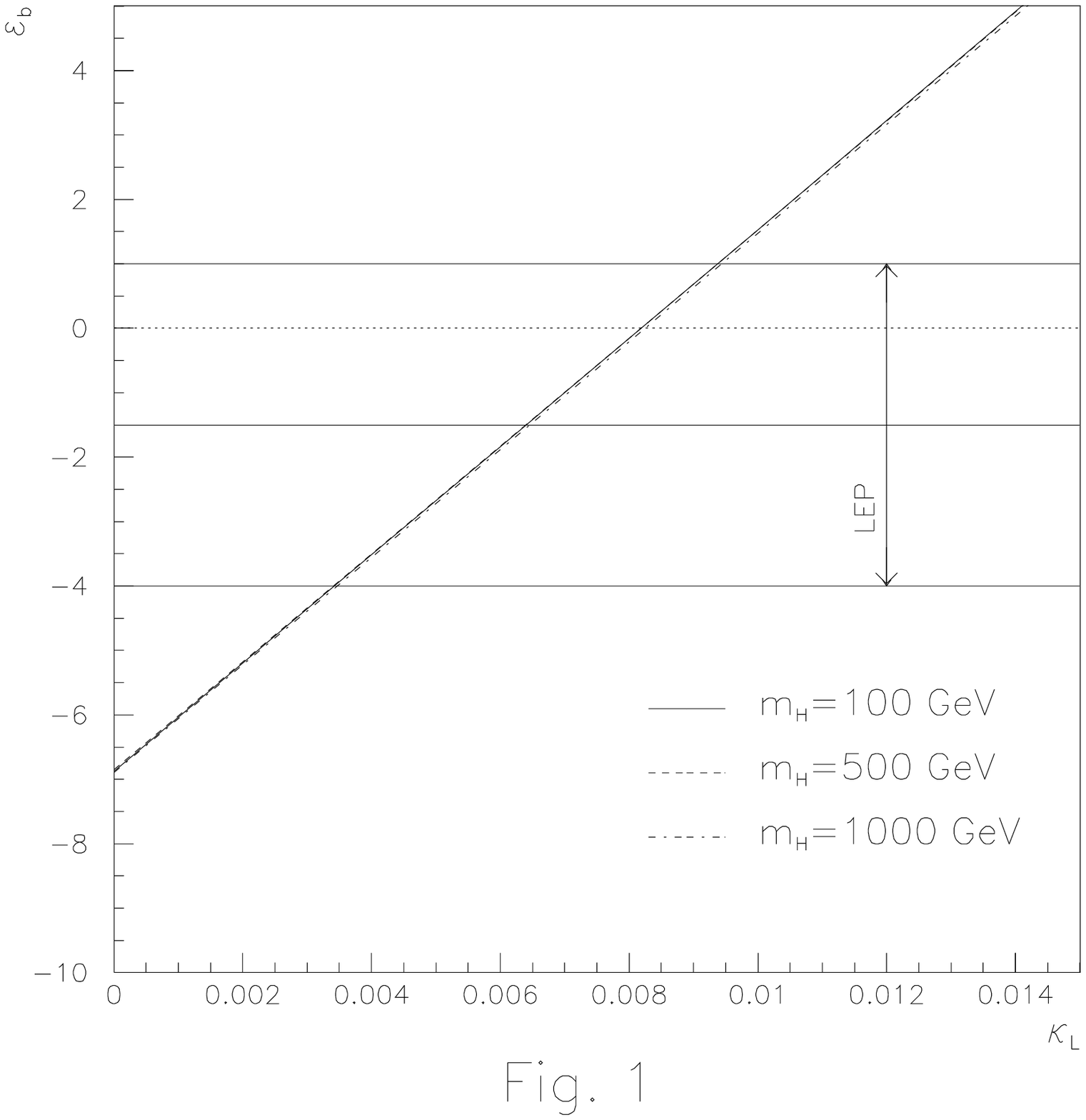}}
\end{figure}

\begin{figure}[th]
\centering
\centerline{\epsfig{file=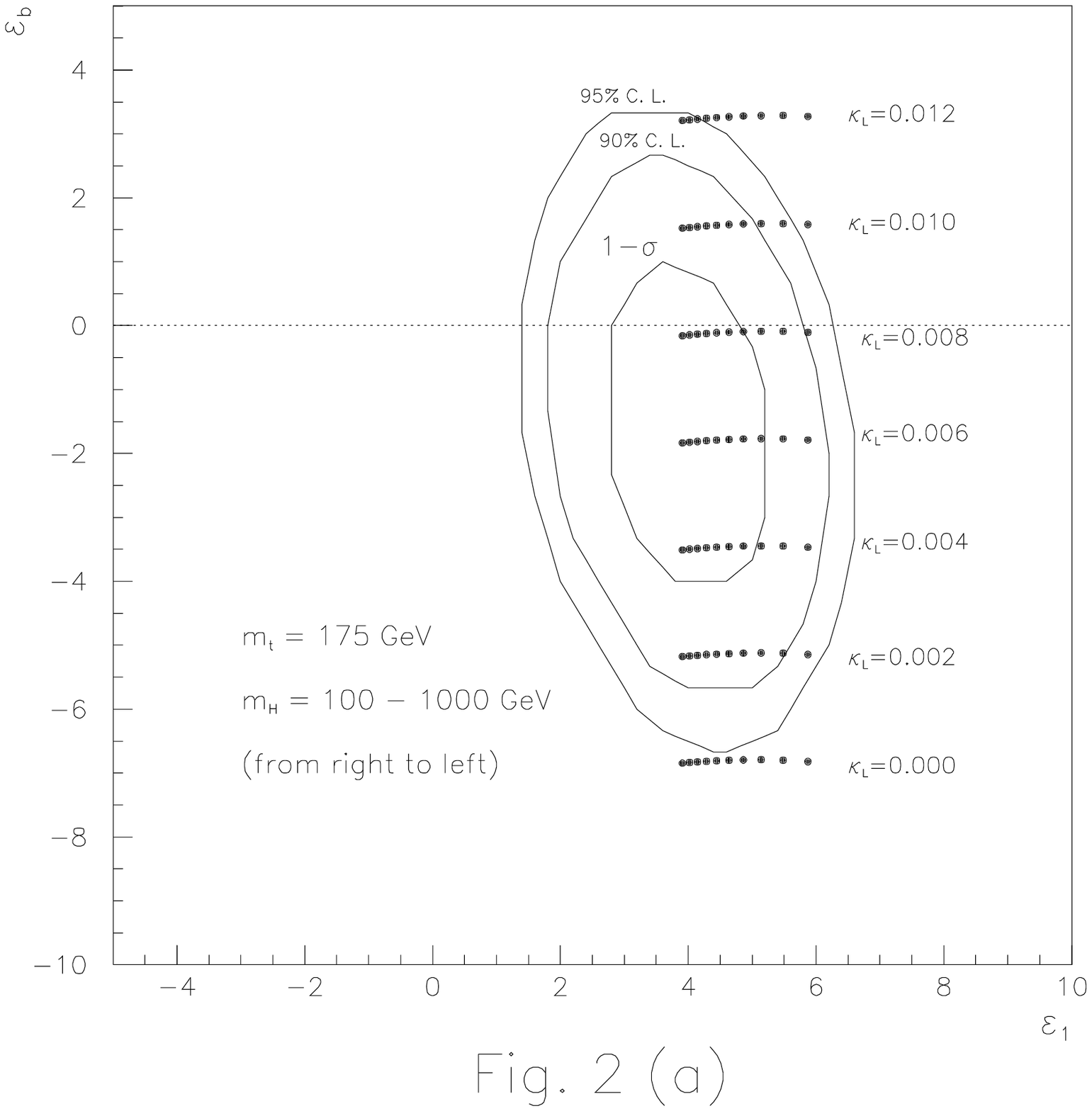}}
\end{figure}

\begin{figure}[th]
\centering
\centerline{\epsfig{file=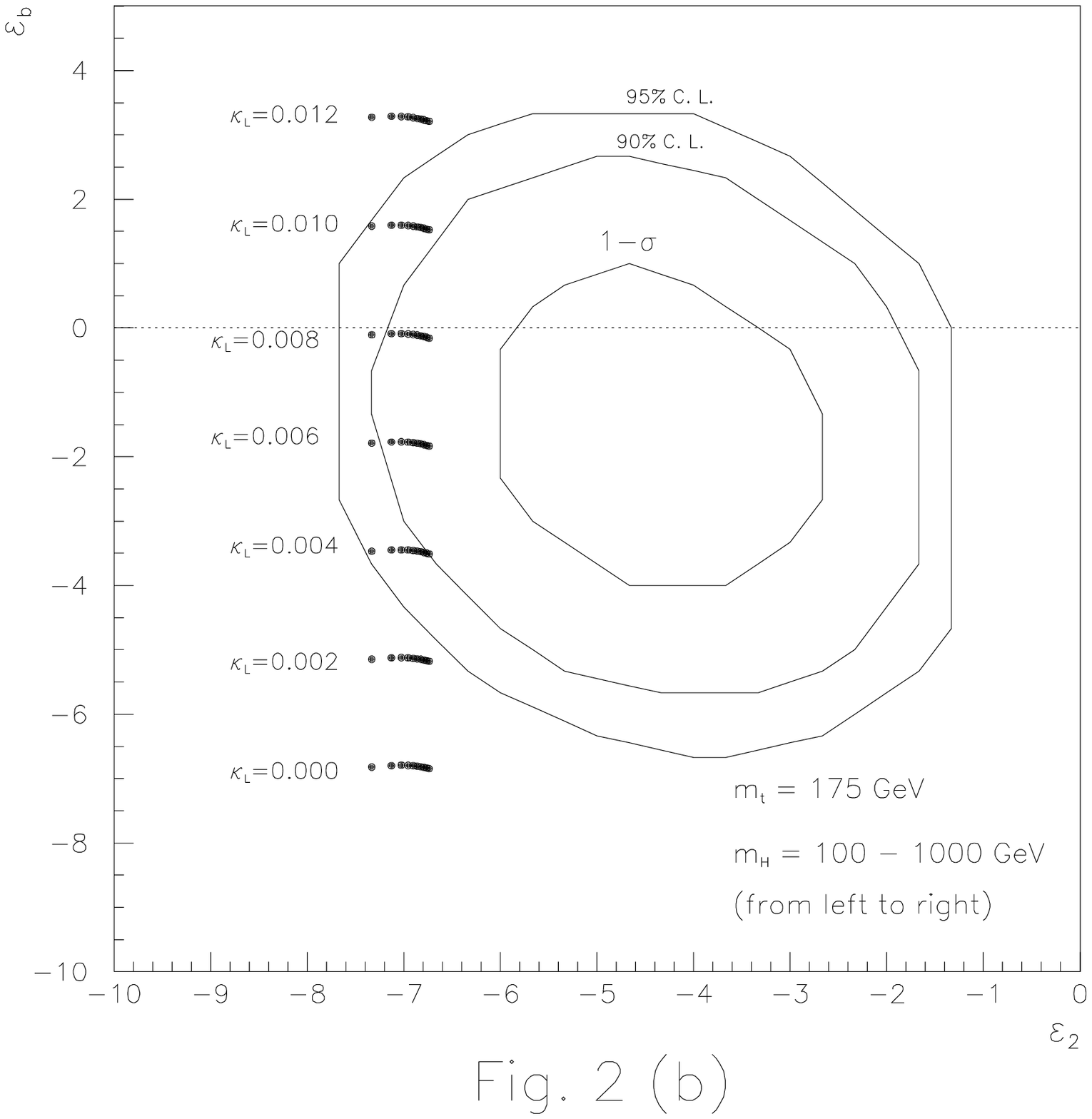}}
\end{figure}

\begin{figure}[th]
\centering
\centerline{\epsfig{file=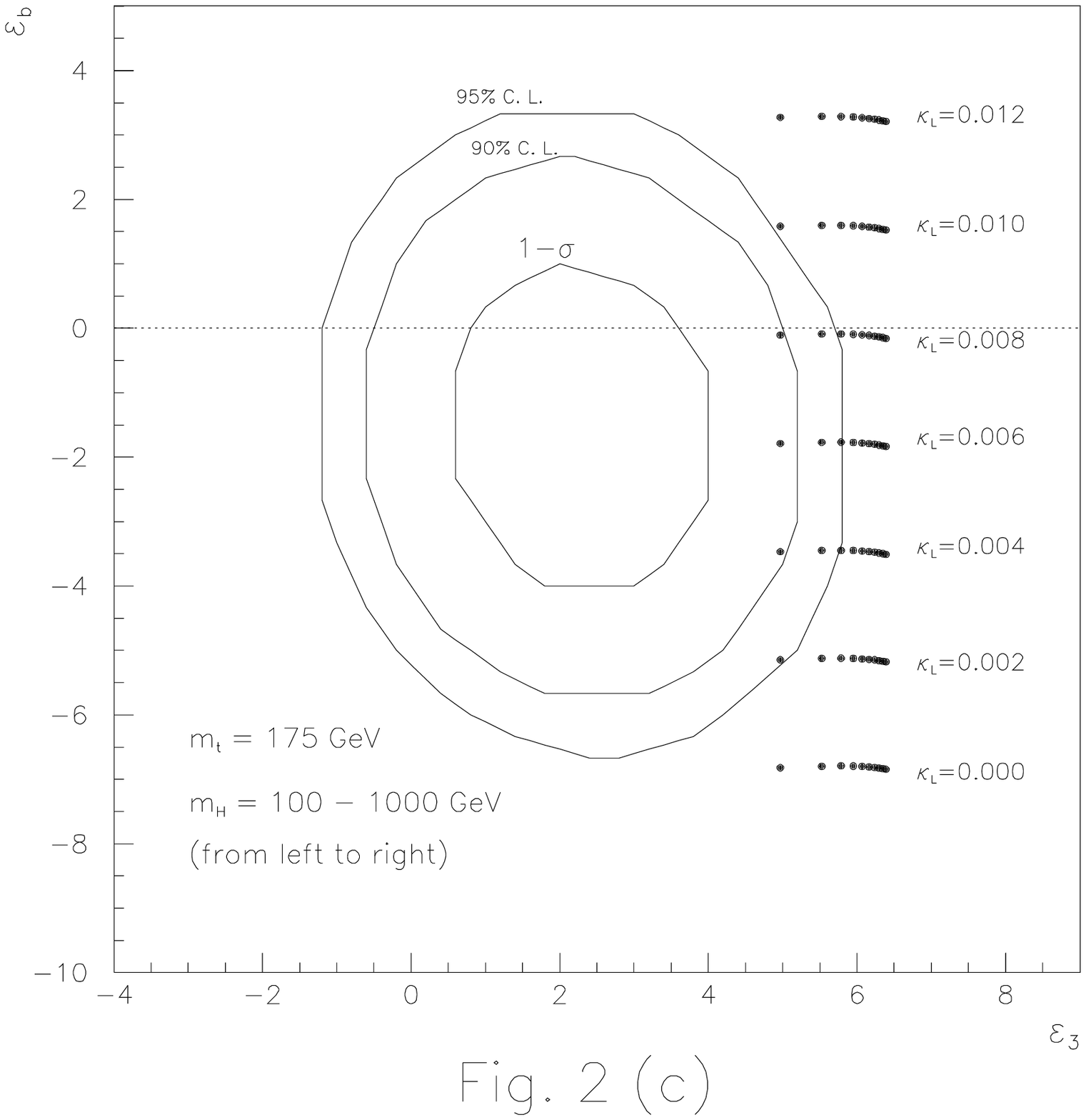}}
\end{figure}

\begin{figure}[th]
\centering
\centerline{\epsfig{file=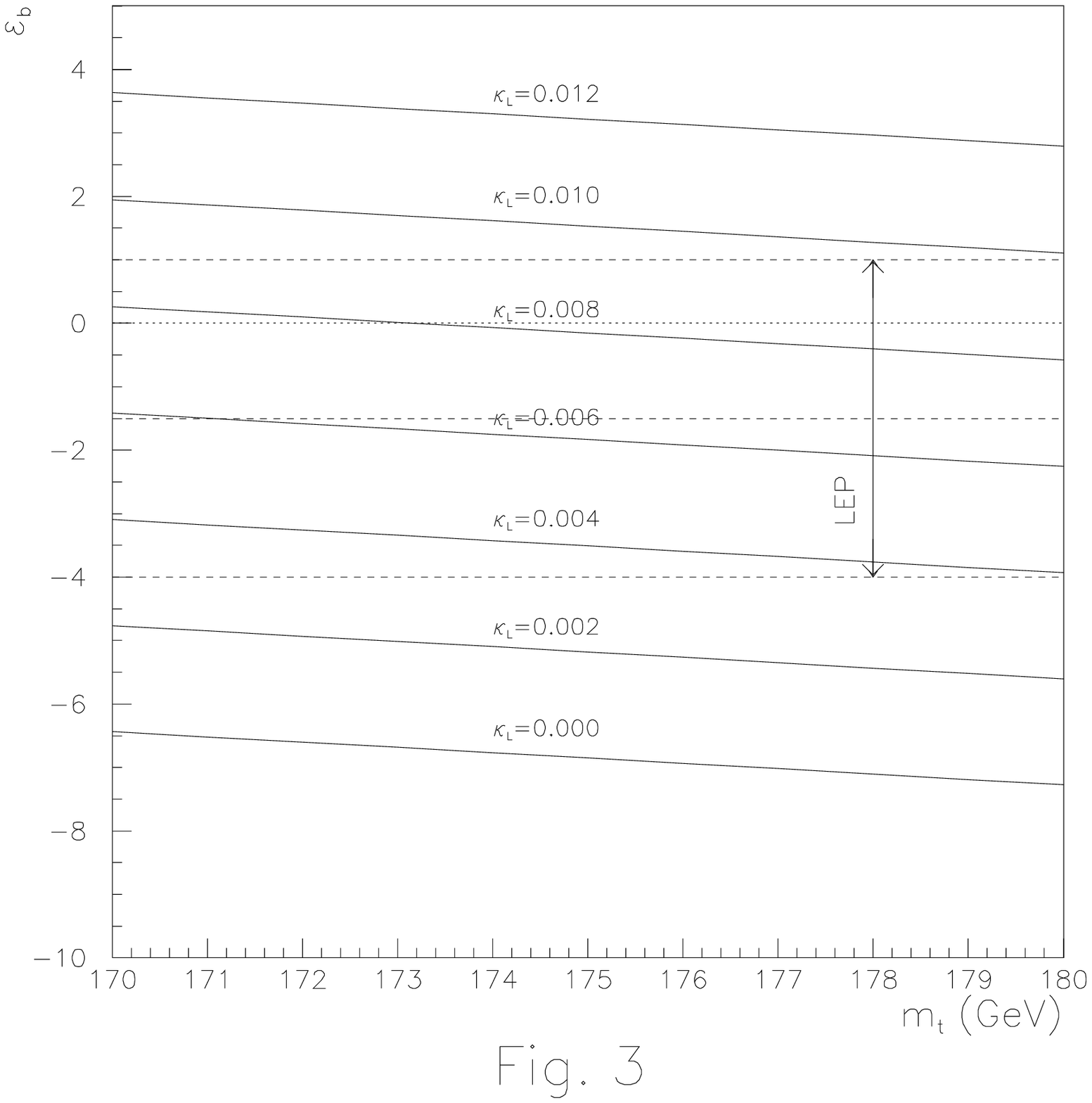}}
\end{figure}

\end{document}